\def\dfrac#1#2{{\displaystyle\frac{#1}{#2}}}

\def\beq{\begin{equation}}
\def\eeq{\end{equation}}
\def\bea{\begin{eqnarray}}
\def\eea{\end{eqnarray}}

\def\non{\nonumber}

\documentclass[a4j,12pt]{article}
\usepackage[dvips]{graphicx}

\begin{document}

\pagestyle{empty}

\thispagestyle{empty}

\begin{center}
{\Large {\bf Determination of $S_{17}$ from Systematic

\vspace{-1mm}

Analysis of $^8$B Coulomb Dissociation

}}

\vspace{5mm}

Kazuyuki Ogata, M. Yahiro$^{\rm A}$, Y. Iseri$^{\rm B}$,
T. Matsumoto, N. Yamashita and\\ M. Kamimura

\vspace{5mm}

{\it Department of Physics, Kyushu University

$^{\rm A}$Department of Physics and Earth Sciences, University of the Ryukyus

$^{\rm B}$Department of Physics, Chiba-Keizai College
}

\end{center}

\begin{abstract}
Systematic analysis of $^8$B Coulomb dissociation with the
Asymptotic Normalization Coefficient (ANC) method is proposed
to determine the astrophysical factor $S_{17}(0)$ accurately.
An important advantage of the analysis is that uncertainties
of the extracted $S_{17}(0)$ coming from the use of the ANC method
can quantitatively be evaluated, in contrast to previous analyses
using the Virtual Photon Theory (VPT).
Calculation of measured spectra in dissociation experiments
is done by means of
the method of Continuum-Discretized Coupled-Channels (CDCC).
From the analysis of $^{58}$Ni($^8$B,$^7$Be$+p$)$^{58}$Ni at 25.8 MeV,
$S_{17}(0)=22.83\pm 0.51({\rm theo})\pm 2.28({\rm expt})$ (eVb)
is obtained; the ANC method turned out to work in this case
within 1\% of error. Preceding systematic analysis of
experimental data at intermediate energies, we propose
hybrid (HY) Coupled-Channels (CC) calculation of $^8$B Coulomb
dissociation, which makes numerical calculation much simple,
retaining its accuracy. The validity of the HY calculation
is tested for $^{58}$Ni($^8$B,$^7$Be$+p$)$^{58}$Ni at 240 MeV.
The ANC method combined with the HY CC calculation is shown to be a
powerful technique to obtain a reliable $S_{17}(0)$.
\end{abstract}

\section{Introduction}
\label{introduction}

The solar neutrino problem is one of the central issues in
the neutrino physics~\cite{Bahcall}.
Nowadays, the neutrino oscillation is assumed
to be the solution of the problem and the focus of
the solar neutrino physics is to determine oscillation parameters:
the mass difference among $\nu_{e}$, $\nu_{\mu}$ and $\nu_{\tau}$,
and their mixing angles~\cite{Bahcall2}.
The astrophysical factor $S_{17}$,
defined by $S_{17}(E)\equiv \sigma_{p\gamma}(E)E\exp [2\pi \eta]$ with
$\sigma_{p\gamma}$ the cross section of the $p$-capture reaction
$^7$Be($p,\gamma$)$^8$B and $\eta$ the Sommerfeld parameter,
plays an essential role in the
investigation of neutrino oscillation,
since the prediction value for the flux of the $^8$B neutrino, which
is intensively being detected on the earth, is proportional to $S_{17}(0)$.
The required accuracy from astrophysics is about 5\% in errors.

Because of difficulties of direct measurements for the $p$-capture reaction
at very low energies, alternative indirect measurements
were proposed: $p$-transfer reactions and
$^8$B Coulomb dissociation
are typical examples of them.
In the former the Asymptotic Normalization Coefficient (ANC)
method~\cite{Xu} is used, carefully evaluating its validity,
while in the latter
the Virtual Photon Theory (VPT) is adopted to extract $S_{17}(0)$;
the use of VPT requires the condition
that the $^8$B is dissociated through its
pure E1 transition, the validity of which is not yet clarified
quantitatively.

In the present paper we propose systematic analysis of
$^8$B Coulomb dissociation by means of the ANC method,
instead of VPT. An important advantage of the analysis is that
one can evaluate the error of $S_{17}(0)$ coming from the use of
the ANC method;
the fluctuation of $S_{17}(0)$, by changing the $^8$B
single-particle wave functions, can be interpreted as
the error of the ANC analysis~\cite{Azhari1,Azhari2,Trache,Ogata}.
For the calculation of $^8$B dissociation cross sections, we use
the method of Continuum-Discretized Coupled-Channels
(CDCC)~\cite{CDCC}, which was proposed and developed by Kyushu group.
CDCC is one of the most accurate methods being applicable to
breakup processes of weakly-bound stable and unstable nuclei.
As a subject of the present analysis, four experiments
of $^8$B Coulomb dissociation done at
RIKEN~\cite{RIKEN}, GSI~\cite{GSI}, MSU~\cite{MSU} and
Notre Dame~\cite{ND} are available. Among them we here take up
the Notre Dame experiment at 25.8 MeV and extract $S_{17}(0)$ by
the CDCC + ANC analysis,
quantitatively evaluating the validity of the use of
the ANC method.

It was shown in Ref.~\cite{MSU} that CDCC can successfully
be applied to the MSU data at 44 MeV/nucleon. However, the CDCC
calculation requires extremely large modelspace; typically
the number of partial waves is 15,000.
Thus, preceding systematic CDCC + ANC analysis of the experimental
data at intermediate energies, we propose hybrid (HY) Coupled-Channels
(CC)
calculation by means of the standard CDCC and the Eikonal-CDCC method
(E-CDCC), which allows one to make efficient and accurate analysis.
E-CDCC describes the center-of-mass (c.m.) motion between the
projectile and the target nucleus by a straight-line,
which is only the essential difference from CDCC.
As a consequence, the resultant E-CDCC equations have a first-order
differential form with no huge angular momenta, hence, one can easily
and safely solve them. Because of the simple straight-line
approximation, results of E-CDCC may deviate from those by
CDCC. One can avoid this problem, however, by constructing
HY scattering amplitude from results of both CDCC and E-CDCC.
This can be done rather straightforwardly,
since the resultant scattering amplitude
by E-CDCC has a very similar form to the quantum-mechanical one,
which is one of the most important features of E-CDCC.
In the latter part of the present paper we show how to perform
the HY calculation and apply it to
$^{58}$Ni($^8$B,$^7$Be$+p$)$^{58}$Ni at 240 MeV.

In Sec.~\ref{ANC+CDCC} we describe the CDCC + ANC analysis for
$^{58}$Ni($^8$B,$^7$Be$+p$)$^{58}$Ni at 25.8 MeV:
the ANC method and CDCC are quickly reviewed in
subsections \ref{ANC+CDCCa} and \ref{ANC+CDCCb}, respectively,
and numerical results and the extracted $S_{17}(0)$
are shown in subsection \ref{ANC+CDCCc}.
In Sec.~\ref{hybrid} the HY calculation for Coulomb dissociation,
with the formalism of E-CDCC, is described
(subsection \ref{hybrida}) and its validity is numerically tested
for $^{58}$Ni($^8$B,$^7$Be$+p$)$^{58}$Ni at 240 MeV
(subsection \ref{hybridb}).
Finally, summary and conclusions are given in Sec.~\ref{summary}.

\section{Systematic analysis of $^8$B Coulomb dissociation}
\label{ANC+CDCC}

In this section we propose CDCC + ANC analysis for
$^8$B Coulomb dissociation to extract $S_{17}(0)$.
First, in subsection~\ref{ANC+CDCCa},
we give a quick review of the ANC method and discuss advantages
of applying it to $^8$B Coulomb dissociation.
Second, calculation of $^8$B breakup cross section by means of
CDCC is briefly described in subsection~\ref{ANC+CDCCb}.
Finally, we show in subsection~\ref{ANC+CDCCc} numerical results for
$^{58}$Ni($^8$B,$^7$Be$+p$)$^{58}$Ni at 25.8 MeV;
the extracted value of $S_{17}(0)$,
with its uncertainties, is given.

\subsection{The Asymptotic Normalization Coefficient method}
\label{ANC+CDCCa}

The ANC method is a powerful
tool to extract $S_{17}(0)$ indirectly.
The essence of the ANC method is that the cross section of the
$^7$Be($p,\gamma$)$^8$B at stellar energies can be determined
accurately if the tail of the $^8$B wave function, described
by the Whittaker function times the ANC, is well determined.
The ANC can be obtained from alternative reactions where peripheral
properties hold well, i.e., only the tail of the $^8$B wave
function has a contribution to observables.

So far the ANC method has been successfully applied
to $p$-transfer reaction such as
$^{10}$Be($^7$Be,$^8$B)$^{9}$Be~\cite{Azhari1},
$^{14}$N($^7$Be,$^8$B)$^{13}$C~\cite{Azhari2},
and $^{7}$Be($d,n$)$^{8}$B~\cite{Ogata}.
Also Trache {\it et al.}~\cite{Trache} showed
the applicability of the ANC method
to one-nucleon breakup reactions;
$S_{17}(0)$ was extracted from systematic analysis of total breakup
cross sections of $^8$B $\longrightarrow$ $^7$Be + $p$ on several
targets at intermediate energies.

In the present paper we apply the ANC method to $^8$B Coulomb dissociation,
where $S_{17}(0)$ has been extracted by using VPT based
on the principle of detailed balance.
In order to use VPT, the previous analyses neglected effects of nuclear
interaction on the $^8$B dissociation, which is not yet well justified.
Additionally, roles of the E2 component, interference with the dominant
E1 part in
particular, need more detailed investigation, although recently
some attempts to
eliminate the E2 contribution from measured spectra have been made.
On the contrary, the ANC analysis proposed here is free from these
problems. We here stress that as an important advantage of the present
analysis, one can evaluate quantitatively
the error of $S_{17}(0)$ by the fluctuation
of the ANC with different $^8$B single-particle potentials.

Comparing with Ref.~\cite{Trache}, in the present ANC analysis
angular distribution and
parallel-momentum distribution of the $^7$Be fragment, instead of
the total breakup cross sections, are investigated,
which is expected to give more accurate value of $S_{17}(0)$.
Moreover, our purpose is to make systematic analysis of $^8$B
dissociation at not only intermediate energies but also quite low
energies. Thus, the breakup process should be described by a
sophisticated reaction theory, beyond the extended Glauber model
used in Ref.~\cite{Trache}.
For that purpose, we use CDCC, which is one of the most accurate methods
to be applicable to $^8$B dissociation.

\subsection{The method of Continuum-Discretized Coupled-Channels}
\label{ANC+CDCCb}

Generally CDCC describes the projectile (c) + target (A) system by a
three-body model as shown in Fig.~1;
%
%
\begin{figure}[htbp]
\begin{center}
\includegraphics[width=70mm,keepaspectratio]{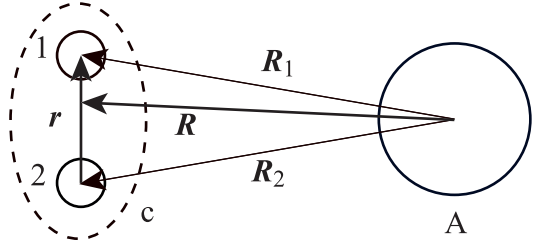}
\end{center}
\vspace*{-5mm}
\caption{
Schematic illustration of the system treated in the present paper.
}
\end{figure}
in the present case c is $^8$B and 1 and 2 denote $^7$Be and $p$,
respectively.
The three-body wave function $\Psi_{JM}$,
corresponding to the total angular momentum $J$ and its projection $M$,
is given in terms of the internal wave functions $\varphi$ of c:
\beq
\Psi_{JM}
=
\sum_{L}
{\cal Y}_{JM}^{\ell_0 L}
\varphi_{0}(r)\frac{\chi_{\ell_0 L J}(P_0,R)}{R}
+
\sum_{\ell L}
{\cal Y}_{JM}^{\ell L}
\int_0^\infty \varphi_{\ell}(k,r)\frac{\chi_{\ell L J}(P,R)}{R} dk;
\label{expand1}
\eeq
\beq
{\cal Y}_{JM}^{\ell L}\equiv
[i^\ell Y_\ell (\Omega_r)\otimes i^L Y_L (\Omega_R)]_{JM},
\eeq
where $\ell$ is the total spin of c and $L$ is the orbital angular
momentum for the relative motion of c and A;
the subscript 0 represents the initial state.
We neglect all intrinsic spins of the constituents and
as a consequence c has only one bound state in the present case.
The first and second terms in the r.h.s. of Eq.~(\ref{expand1}) correspond to
the bound and scattering states of c, respectively. In the latter
the relative momentum $P$ between c and A is related to the internal
one $k$ of c through the total-energy conservation.

In CDCC the summation over $\ell$ and integration over $k$ are
truncated at certain values $\ell_{\rm max}$ and $k_{\rm max}$,
respectively. For the latter, furthermore, we divide the $k$ continuum
into $N$ {\it bin}-states, each of which is expressed by a discrete
state $\hat{\varphi}_{i \ell}$ with $i$ denote a certain region of $k$,
i.e., $k_{i-1} \le k < k_i$.
After truncation and discretization, $\Psi_{JM}$ is
approximately expressed by $\{ \hat{\varphi}_{i \ell} \}$ with
finite number of channels:
\beq
\Psi_{JM}^{\rm CDCC}
=
\sum_{L}
{\cal Y}_{JM}^{\ell_0 L}
\varphi_{0}(r)\frac{\chi_{\ell_0 L J}(P_0,R)}{R}
+
\sum_{\ell=0}^{\ell_{\rm max}}
\sum_{i=1}^{N}
\sum_{L}
{\cal Y}_{JM}^{\ell L}
\hat{\varphi}_{i \ell}(r)\frac{\hat{\chi}_{\gamma}(\hat{P}_i,R)}{R}
\label{expand2}
\eeq
with $\gamma=\{i,\ell,L,J\}$.
The $\hat{P_i}$ and $\hat{\chi}_\gamma$ are
the discretized $P$ and $\chi_{\ell L J}$, respectively, corresponding
to the $i$th bin state $\hat{\varphi}_{i \ell}$.

Inserting $\Psi_{JM}^{\rm CDCC}$ into a three-body
Schr\"{o}dinger equation, one obtains the following (CC) equations:
\beq
\left[\frac{d^{2}}{dR^{2}} + \hat{P}_{i}^{2} - \frac{L(L+1)}{R^{2}}
 - \frac{2 \mu}{\hbar^{2}} V_{\gamma \gamma}(R)\right]
\hat{\chi}_{\gamma}(\hat{P}_i,R)
 = \sum_{\gamma^{'}\neq\gamma} \frac{2 \mu}{\hbar^{2}}
   V_{\gamma \gamma ^{'}}(R) \hat{\chi}_{\gamma^{'}} (\hat{P}_{i^{'}},R)
\label{eqs:CDCC}
\eeq
for all $\gamma$ including the initial state, where
$\mu$ is the reduced mass
of the c + A system and $V_{\gamma \gamma'}$ is the form factor
defined by
\begin{equation}
  V_{\gamma \gamma^{'}}(R)   =
     \langle {\cal Y}_{JM}^{\ell L}\hat {\varphi}_{i\ell} (r)
     | U | {\cal Y}_{JM}^{\ell^{'} L^{'}}\hat {\varphi}_{i^{'}\ell^{'}} (r)
     \rangle_{{\bf r}, \Omega_R},
\label{coupling}
\end{equation}
with $U$ the sum of the interactions between A and individual constituents
of c.
The CDCC equations (\ref{eqs:CDCC}) are solved with the
asymptotic boundary condition:
\begin{equation}
 \hat{\chi}_{\gamma}(\hat{P}_i,R) \sim u_{L}^{(-)} (\hat{P}_{i},R)
\delta_{\gamma , \gamma_0} - \sqrt{\hat{P}_i/\hat{P}_{0}}
\hat{S}_{\gamma , \gamma_0} u_{L}^{(+)} ( \hat{P}_{i} , R),
 \end{equation}
where
$u_{L}^{(-)}$ and $u_{L}^{(+)}$
are incoming and outgoing Coulomb wave functions.
Thus one obtains the $S$-matrix elements
$\hat{S}_{\gamma , \gamma_0}$, from which any observables, in principle,
can be calculated;
we followed Ref.~\cite{Iseri} to calculate the distribution of $^7$Be
fragment from $^8$B.

CDCC treats breakup chanels of a projectile
explicitly, including all higher-order
terms of both Coulomb and nuclear coupling-potentials, which
gives very accurate description of dissociation processes
in a framework of three-body reaction dynamics.
Detailed formalism and theoretical foundation of CDCC can be found
in Refs.~\cite{CDCC,Piya,CDCC-foundation}.

\subsection{Numerical results and the extracted $S_{17}(0)$}
\label{ANC+CDCCc}

We here take up the $^8$B dissociation by $^{58}$Ni at 25.8 MeV
(3.2 MeV/nucleon) measured at Notre Dame~\cite{ND}, for which VPT was found
to fail to reproduce the data~\cite{EB}. The extended Glauber model,
used in Ref.~\cite{Trache}, is
also expected not to work well because of the low incident energy.
Thus, the Notre Dame data is a good subject of our CDCC + ANC analysis.

Parameters of the modelspace taken in the CDCC calculation are as follows.
The number of bin-states of $^8$B
is 32 for s-state and 16 for p-, d- and f-states.
We neglected the intrinsic spins of $p$, $^7$Be and $^{58}$Ni
as mentioned in the previous subsection.
The maximum excitation energy of $^8$B is 10 MeV,
$r_{\rm max}$ ($R_{\rm max}$) is 100 fm (500 fm) and
$J_{\rm max}$ is 1000.
For nuclear interactions of $p$-$^{58}$Ni and $^7$Be-$^{58}$Ni
we used the parameter sets of
Becchetti and Greenlees~\cite{BG} and
Moroz {\it et al}.~\cite{Moroz}, respectively.

%
%
\begin{figure}[htbp]
\begin{center}
\includegraphics[width=140mm,keepaspectratio]{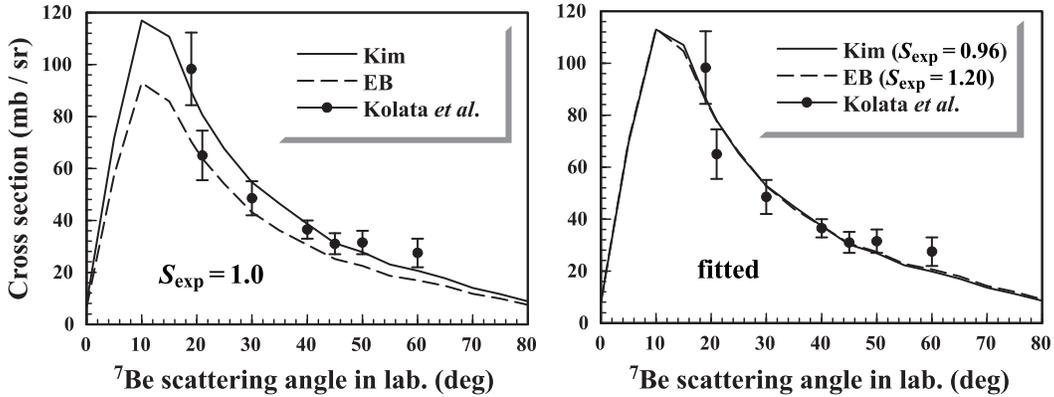}
\end{center}
\vspace*{-3mm}
\caption{
Angular distribution of
the $^7$Be fragment in the laboratory frame.
The solid and dashed lines represent the results of CDCC
calculation with the parameter set of Kim and Esbensen-Bertsch (EB),
respectively,
for $^8$B single particle potential.
Results in the left panel correspond to $S_{\rm exp}=1$
and those with appropriate values of $S_{\rm exp}$, i.e.,
0.96 for Kim and 1.20 for EB, are shown in the right panel.
The experimental data are taken from Ref.~\protect\cite{ND}.
}
\end{figure}
In Fig.~2 we show the results of the angular distribution of
$^7$Be fragment, integrated over scattering angles of $p$ and
excitation energies of the $^7$Be + $p$ system.
In the left panel the results with the
$^8$B wave functions by Kim {\it et al.}~\cite{Kim} (solid line)
and Esbensen and Bertsch~\cite{EB} (dashed line), with the spectroscopic
factor $S_{\rm exp}$ equal to unity, are shown. After $\chi^2$
fitting, one obtains the results in the right panel; one sees
that both calculations very well reproduce the experimental
data. The resultant $S_{\rm exp}$ is 0.96 and 1.20 with
the $^8$B wave functions by Kim and Esbensen-Bertsch, respectively,
showing quite strong dependence on $^8$B models.
In contrast to that, the ANC $C$ calculated by
$C=S_{\rm exp}^{1/2}b$
with $b$ the single-particle ANC, is found to be
almost independent of the choice
of $^8$B wave functions, i.e., $C=0.59\pm0.004$ (fm$^{-1/2}$).
Thus, one can conclude
that the ANC method works in the present case within 1\% of error.

Following Ref.~\cite{Xu} we obtained the following result:
\[
S_{17}(0)=22.83\pm 0.17({\rm ANC})\pm 0.34({\rm CDCC})\pm 2.28({\rm expt})
\;\;({\rm eVb}),
\]
where the uncertainties from the choice of the modelspace of
CDCC calculation (1.5\%) and the systematic error of
the experimental data (10\%) are also included.
Although the quite large experimental error prevents one from
determining $S_{17}(0)$ with the required accuracy (5\%),
the CDCC + ANC method turned out to be a powerful technique
to determine $S_{17}(0)$ with small theoretical uncertainties.
More careful analysis in terms of the charge distribution of
$^7$Be, nuclear optical potentials and roles of the intrinsic
spins of the constituents, are being made and more reliable $S_{17}(0)$
will be reported in a forthcoming paper.

\section{Hybrid calculation for Coulomb dissociation}
\label{hybrid}

In Sec.~\ref{ANC+CDCC} we showed that the CDCC + ANC analysis
for the $^8$B dissociation at 25.8 MeV gives $S_{17}(0)$
with good accuracy, being free from rather ambiguous assumptions
made in the previous analyses using VPT.
In Ref.~\cite{Trache} it was shown that
the ANC method works well for
one-nucleon breakup reactions at intermediate energies.
Also CDCC turned out to almost perfectly
reproduce the
parallel-momentum distribution of $^7$Be fragment from
$^{208}$Pb($^8$B,$^7$Be$+p$)$^{208}$Pb at 44 MeV/nucleon~\cite{MSU}.
Thus, it is expected that accurate determination
of $S_{17}(0)$ can be done by the CDCC + ANC analysis of
$^8$B Coulomb dissociation measured at RIKEN~\cite{RIKEN},
GSI~\cite{GSI} and MSU~\cite{MSU}.

From a practical point of view, however, CDCC calculation
including long-ranged Coulomb coupling-potentials requires
extremely large modelspace rather difficult to handle;
typically the number of partial waves is 15,000 for
the MSU data~\cite{MSU}.
Although interpolation technique for angular momentum reduces the
number of CC equations to be solved in terms of $J$,
those with huge angular momenta are rather
unstable and careful treatment is necessary.
In this sense, it seems almost impossible to apply CDCC to the GSI
data at 250 MeV/nucleon, where $J_{\rm max}$ is expected
to exceed 100,000.

On the contrary, semi-classical approaches, expected to work quite
well at intermediate energies, are free from any problems concerned with
huge angular momenta. The accuracy of such semi-classical calculations,
however, is difficult to be evaluated quantitatively,
although one may naively estimate
the error is only less than about 10\% or so. It should be noted
that our goal is to determine $S_{17}(0)$ with more than 95\% accuracy,
which requires definite estimation of the error of the calculation.

In this section we propose HY calculation of $^8$B dissociation
at intermediate energies, constructing HY scattering amplitude
($T$ matrix) from partial amplitudes with quantum-mechanical
(QM) and eikonal (EK) CC calculations; for the latter we use
a new version of CDCC, that is, the Eikonal-CDCC method (E-CDCC).
The formalism of E-CDCC and calculation of the HY amplitude
are described in subsection \ref{hybrida} and
the validity of the hybrid calculation is tested for
$^{58}$Ni($^8$B,$^7$Be$+p$)$^{58}$Ni at 240 MeV in subsection \ref{hybridb}.

\subsection{The Eikonal-CDCC method and construction of hybrid
scattering amplitude}
\label{hybrida}

We start with the expansion of the total wave function $\Psi$:
\beq
\Psi({\bf R},{\bf r})
=
\sum_{i \ell m}
\Phi_{i,\ell m}({\bf r})
e^{-i(m-m_0)\phi_R}
\chi_{i \ell m}(R,\theta_R),
\label{psinew2}
\eeq
where $m$ is the projection of $\ell$ on the $z$-axis taken to be parallel
to the incident beam;
$\Phi_{i,\ell m}$ is the discretized internal-wave-function of c,
calculated just in the same way as in the standard CDCC.
The symbol {\lq\lq} $\hat{}$ '' used in subsection \ref{ANC+CDCCb},
which denotes a discretized quantity, is omitted here for simplicity.

We make the following EK approximation:
\beq
\chi_{c}(R,\theta_R)
\approx
\psi_{c}(b,z)
\dfrac{1}{(2\pi)^{3/2}}
e^{i{\bf K}_c(b) \cdot {\bf R}},
\label{eikonal}
\eeq
where $c$ denotes channels \{$i$, $\ell$, $m$\} together and
the wave number $K_c$ is defined by
\beq
\dfrac{\hbar^2}{2\mu}K_c^2(b)=
E-\epsilon_{i,\ell}-\dfrac{\hbar^2}{2\mu}
\dfrac{(m-m_0)^2}{b^2}
\label{capk}
\eeq
with $b$ the impact parameter;
the direction of ${\bf K}_c$ is assumed to be parallel to the $z$-axis.

Inserting Eqs.~(\ref{psinew2}) and (\ref{eikonal}) into a three-body
Schr\"{o}dinger equation and neglecting
the second order derivative of $\psi_{c}$, one can obtain
the following E-CDCC equations:
\beq
\dfrac{i\hbar^2}{\mu}
K_c^{(b)}
\dfrac{d}{d z}\psi_{c}^{(b)}(z)
=
\sum_{c'}
{\cal F}^{(b)}_{cc'}(z)
\;
\psi_{c'}^{(b)}(z)
e^{i\left(K_{c'}^{(b)}-K_c^{(b)}\right) z}
\label{cceq4}
\eeq
for all $c$ including $c_0$, with
${\cal F}^{(b)}_{cc'}(z)=
\langle \Phi_{c} ({\bf r})
| U | \Phi_{c'} ({\bf r})
\rangle_{\bf r}\exp[i(m'-m)\phi_R]$.
We put $b$ in a superscript since
it is not a dynamical variable but an input parameter.
Equations (\ref{cceq4}) are solved with the boundary condition
$\psi_{c}^{(b)}(-\infty)=\delta_{c0}$.
Since the E-CDCC equations are first-order differential ones and
contain no coefficients with huge angular momenta, they can easily
and safely be solved.

Using the solutions of Eq.~(\ref{cceq4}), the
scattering amplitude with E-CDCC is given by
\beq
f_{c0}^{\rm E}
=
-\dfrac{\mu}{2\pi\hbar^2}
\int
\sum_{c'}
{\cal F}^{(b)}_{cc'}(z)
\,
e^{-i(m-m_0)\phi_R}
e^{i ({\bf K_{c'}^{(b)}}-{\bf K_c'^{(b)}})\cdot {\bf R}}
\,
\psi_{c'}^{(b)}(z)
d{\bf R}.
\label{f}
\eeq
Making use of the following forward-scattering approximation:
\beq
({\bf K_{c'}^{(b)}}-{\bf K_c'^{(b)}})\cdot {\bf R}
\approx
-K_c^{(b)}\theta_f b\cos{\phi_R}+(K_{c'}^{(b)}-K_c^{(b)})z,
\label{C}
\eeq
one obtains
\beq
f_{i \ell m,i_0 \ell_0 m_0}^{\rm E}
=
\dfrac{1}{2\pi i}
\int
\!\!
\int
K_{i \ell m}^{(b)}
e^{-i(m-m_0)\phi_R}
e^{-iK_{i \ell m}^{(b)} \theta_f b\cos{\phi_R}}
\left(
{\cal S}_{i \ell m,i_0 \ell_0 m_0}^{(b)}
\!\!-
\delta_{i i_0}\delta_{\ell \ell_0}\delta_{m m_0}
\right)
bdbd\phi_R,\\
\label{f4}
\eeq
where
the EK $S$-matrix elements are defined by
${\cal S}_{i \ell m,i_0 \ell_0 m_0}^{(b)}\equiv
\psi_{i \ell m}^{(b)}(\infty)$.

We then {\it discretize} $f^{\rm E}$:
\bea
f_{i \ell m,i_0 \ell_0 m_0}^{\rm E}
&=&
\dfrac{1}{2\pi i}
\sum_L
K_{i \ell m}^{(b_L^{\rm mid})}
\left[
\int
e^{-i(m-m_0)\phi_R}
e^{-iK_{i \ell m}^{(b_L^{\rm mid})}\theta_f b_L\cos{\phi_R}}
d\phi_R
\right]
\non \\
& &
\times{}
\left(
{\cal S}_{i \ell m,i_0 \ell_0 m_0}^{(b_L^{\rm mid})}
-
\delta_{i i_0}\delta_{\ell \ell_0}\delta_{m m_0}
\right)
\int_{b_L^{\rm min}}^{b_L^{\rm max}}
bdb,
\label{f6}
\eea
where
$b_L^{\rm min}$, $b_L^{\rm max}$ and $b_L^{\rm mid}$
are defined through
$K_{i \ell m}^{(b_L^{\rm min})}b_L^{\rm min}= L$,
$K_{i \ell m}^{(b_L^{\rm max})}b_L^{\rm max}= L+1$
and $K_{i \ell m}^{(b_L^{\rm mid})}b_L^{\rm mid}= L+1/2$,
respectively. In deriving Eq.~(\ref{f6}) we
neglected the $b$-dependence of
$K_{i \ell m}^{(b)}$,
$\exp[-iK_{i \ell m}^{(b)}\theta_f b\cos{\phi_R}]$
and
${\cal S}_{i \ell m,i_0 \ell_0 m_0}^{(b)}$
within a small size of $b$ corresponding to each $L$.
After manipulation one can obtain
\beq
f_{i \ell m,i_0 \ell_0 m_0}^{\rm E}
\!\!\approx
\dfrac{2\pi}{i K_0}
\sum_L
\dfrac{K_0}{K_{i \ell m}^{(b_L^{\rm mid})}}
\sqrt{\dfrac{2L+1}{4\pi}}
i^{(m-m_0)}
Y_{L \,m\!-m_0}(\hat{\bf K}')
\left(
{\cal S}_{i \ell m,i_0 \ell_0 m_0}^{(b_L^{\rm mid})}
\!\!-
\delta_{i i_0}\delta_{\ell \ell_0}\delta_{m m_0}
\right),
\non \\
\label{f7}
\eeq
which has a similar form to that of the standard CDCC:
\bea
f^{\rm Q}_{i \ell m,i_0 \ell_0 m_0}
\!\!\!\!\!&=&\!\!\!
\dfrac{2\pi}{iK_0}
\sum_{L}
\sum_{J=|L-\ell|}^{L-\ell}
\sum_{L_0=|J-\ell_0|}^{J-\ell_0}
\sqrt{\dfrac{2L_0+1}{4\pi}}
(L_0 0 \ell_0 m_0 | J m_0)
(L \;m_0\!\!-\!m \;\ell m | J m_0)
\non \\
\!\!\!& &\!\!\!
\hspace{16mm}
{}\times
(S_{i L \ell,i_0 L_0 \ell_0}^{J}-\delta_{i i_0}
\delta_{L L_0}\delta_{\ell \ell_0})
(-)^{m-m_0}Y_{L \,m\!-m_0}(\hat{\bf K}').
\label{fq}
\eea

The construction of the HY scattering amplitude $f^{\rm H}$
can be done by:
\beq
f_{i \ell m,i_0 \ell_0 m_0}^{\rm H}
\equiv
\sum_{L\le L_{\rm C}}
f_L^{\rm Q}
+
\sum_{L > L_{\rm C}}
f_L^{\rm E},
\label{fh}
\eeq
where $f_L^{\rm Q}$ ($f_L^{\rm E}$) is the $L$-component of
$f^{\rm Q}$ ($f^{\rm E}$) and
$L_{\rm C}$ represents the connecting point between
the QM and EK calculations, which is chosen
so that $f_L^{\rm E}$ coincides with $f_L^{\rm Q}$ for
$L>L_{\rm C}$.
One sees that Eq.~(\ref{fh}) includes all QM
effects necessary through $f_L^{\rm Q}$, and also interference
between the lower and higher $L$-regions.
It should be noted that derivation of Eq.~(\ref{f7}), which leads one
to Eq.~(\ref{fh}) rather straightforwardly, is one of the most important
features of E-CDCC. Actually, the present EK calculation is very simple,
i.e., E-CDCC equations (\ref{cceq4})
contain no correction terms to the straight-line
approximation. However, this is not a defect but a merit of E-CDCC, since
such a simplest calculation is enough to describe scattering
processes, if combined with the result of the QM calculation
taking an appropriate $L_{\rm C}$.

In the above formulation we neglected the Coulomb distortion.
In order to include it, we use
$
\chi_{c}(R,\theta_R)
\approx
\psi_{c}(b,z)
(2\pi)^{-3/2}
\phi_{c}^{\rm C}(b,z)
\label{eikonal2}
$
instead of Eq.~(\ref{eikonal})~\cite{Kawai}, where $\phi_{c}^{\rm C}$
is the Coulomb wave function.
The formulation of $f^{\rm E}$ can then be done just in the same way
as above.

\subsection{Numerical test for the hybrid calculation}
\label{hybridb}

In order to see the validity of the HY calculation with
CDCC and E-CDCC, we analyze $^{58}$Ni($^8$B,$^7$Be$+p$)$^{58}$Ni
at 240 MeV.
The number of bin-states of $^8$B
is 16, 8 and 8 for s-, p- and d-states, respectively, and
$L_{\rm max}$ is 4000.
As for the $^8$B wave function, the parameter set by
Kim {\it et al.}~\cite{Kim} was adopted.
For nuclear interaction between $^7$Be and $^{58}$Ni
we used the global potential for $^7$Li scattering
by Cook {\it et al}.~\cite{Cook}.
Other parameters are taken just the same as in subsection \ref{ANC+CDCCc}.
%
%
\begin{figure}[htbp]
\begin{center}
\includegraphics[width=140mm,keepaspectratio]{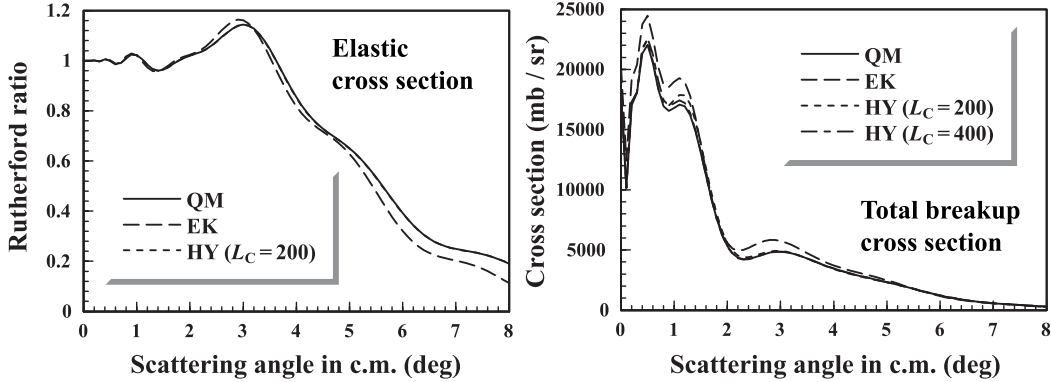}
\end{center}
\vspace*{-3mm}
\caption{
Angular distribution of
the elastic (left panel) and total breakup (right panel)
cross sections for
$^{58}$Ni($^8$B,$^7$Be$+p$)$^{58}$Ni at 240 MeV.
The solid and dashed lines
show, respectively, the results of the QM
and EK calculations.
The dotted (dash-dotted) line represents the HY result
with $L_{\rm C}=200$ (400).
}
\end{figure}

In the left and right panels in Fig.~3 we show
the elastic cross section (Rutherford ratio) and the total breakup one,
respectively,
as a function of scattering angle in the center-of mass (c.m.)
frame.
The solid, dashed and dotted lines
represent the results of the QM,
EK and HY calculations, where $L_{\rm C}$ is
taken to be 4000, 0 and 200, respectively.
In the right panel the result of the HY calculation with $L_{\rm C}=400$
is also shown by the dash-dotted line.
The agreement between the QM and HY calculations with an appropriate
value of $L_{\rm C}$ for the latter,
namely, 200 (400) for elastic (breakup) cross
section, is excellent;
the error is only less than 1\%.
One also sees that difference between the EK and QM results is appreciable.
Since the present EK calculation is quite simple, as mentioned in the previous
subsection, this does not directly show the fail of EK approximation.
However, it seems quite difficult for EK calculation to
obtain {\lq\lq}perfect'' agreement with the result of the QM one.
On the contrary, the HY calculation turned out to be applicable to
analyses of $^8$B dissociation to extract $S_{17}(0)$,
where very high accuracy is required.
%
%
\begin{figure}[btbp]
\begin{center}
\includegraphics[width=140mm,keepaspectratio]{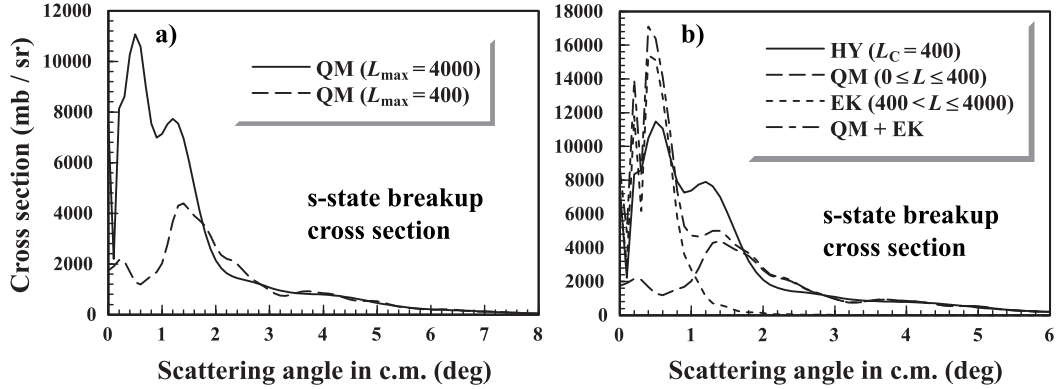}
\end{center}
\vspace*{-3mm}
\caption{
a) QM results for the s-state breakup cross
section with $L_{\rm max}=4000$ (solid line)
and 400 (dashed line).
b) The s-state breakup cross sections by
the QM calculation with $0 \le L \le 400$ (dashed line)
and the EK calculation with $400< L \le 4000$ (dotted line).
The dash-dotted line is the incoherent sum of the
dashed and dotted lines and the solid line is the
HY result with $L_{\rm C}=400$, namely, the coherent sum of the two.
}
\end{figure}

We show in the left panel of Fig.~4 the s-state breakup cross sections
by the QM calculation; the solid and dashed lines correspond to the calculation
with $L_{\rm max}=4000$ and 400, respectively.
One sees big difference between the two, which shows that
the partial scattering amplitudes for larger $L$ indeed has
an essential contribution to the breakup cross section.
In the right panel we show the s-state breakup cross sections by
the QM calculation with $0 \le L \le 400$ (dashed line)
and the EK calculation with $400< L \le 4000$ (dotted lines).
The dash-dotted line is the incoherent sum of the two,
which deviates from the HY result
shown by the solid line. Thus, one sees that
the essence of the present HY calculation is the construction of the HY
scattering amplitude not the HY cross section.

We show in Fig.~5 the p-state breakup cross sections by
the QM calculation with (dashed line) and without (solid line)
adiabatic (AD) approximation.
One sees that the AD approximation increases the breakup cross
section about 10\% at forward angles. The oscillation shown by
the dashed line seems to indicate the AD calculation is not
valid in the present case, probably for higher partial waves.
We found just the same features in the EK calculation.
%
%
\begin{figure}[htbp]
\begin{center}
\begin{minipage}[t]{.45\textwidth}
\includegraphics[width=65mm,keepaspectratio]{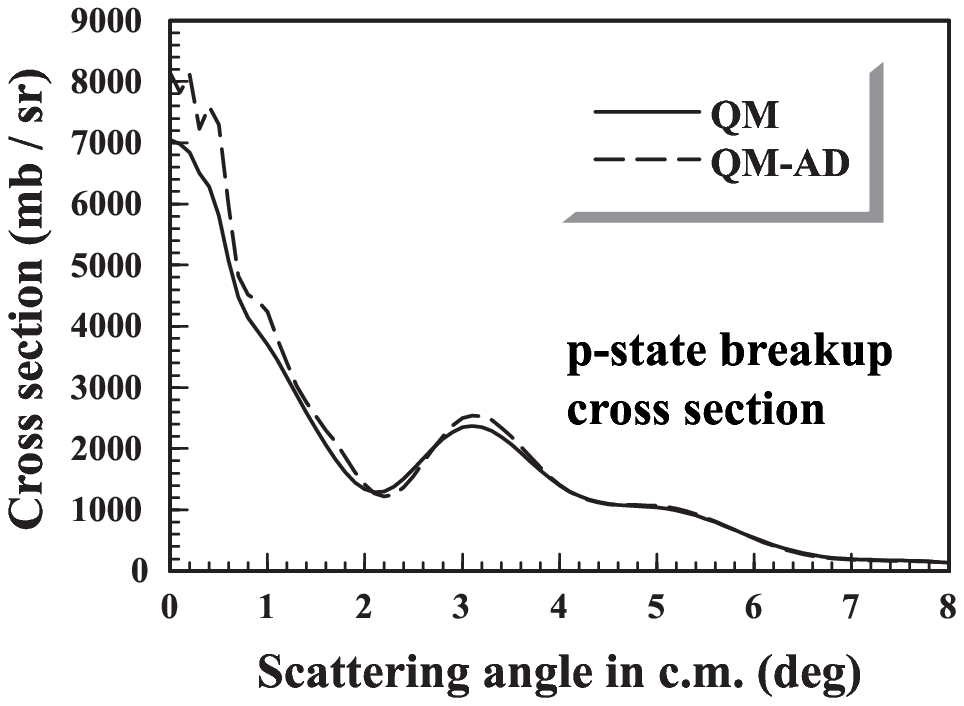}
\caption{
The p-state breakup cross sections by the QM calculation
with (dashed line)
and without (solid line) AD approximation.
}
\end{minipage}
\begin{minipage}[t]{.2\textwidth}
\end{minipage}
\begin{minipage}[t]{.45\textwidth}
\includegraphics[width=70mm,keepaspectratio]{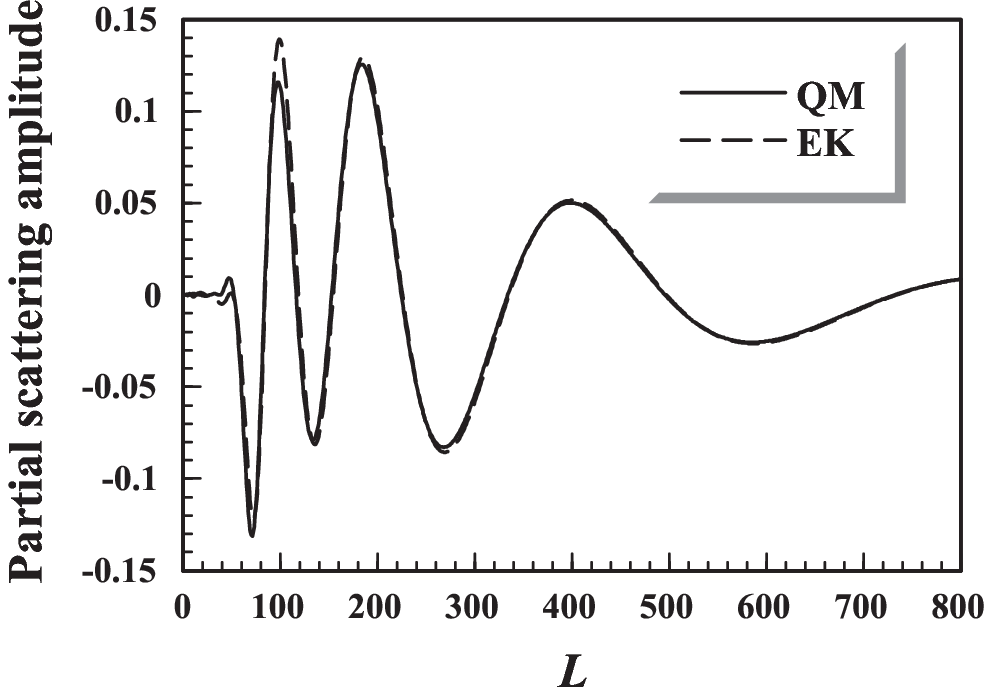}
\caption{
Comparison between $f_L^{\rm Q}$ (solid line)
and $f_L^{\rm E}$ (dashed line) for the
\{$\ell$, $m$\} $=$ \{$\ell_0$, $m_0$\} $=$ \{1, 0\}
component.
}
\end{minipage}
\end{center}
\end{figure}

Comparison between $f_L^{\rm Q}$ (solid line)
and $f_L^{\rm E}$ (dashed line) in Eq.~(\ref{fh}),
corresponding to the
\{$\ell$, $m$\} $=$ \{$\ell_0$, $m_0$\} $=$ \{1, 0\} component,
is made in Fig.~6.
One sees that the difference between $f_L^{\rm Q}$ and $f_L^{\rm E}$
is appreciable for smaller $L$, around 100 in particular,
and as the larger $L$ becomes, the better agreement is obtained.
For $L\ge 400$, the difference is not visible, which is consistent
with the result shown in the right panel of Fig.~3.

Thus, the HY calculation of $^8$B Coulomb dissociation turned out
to allow one to make efficient and accurate analysis. The method is
expected to be applicable to the experimental data measured at not
only RIKEN and MSU (at several tens of MeV/nucleon)
but also GSI (at 250 MeV/nucleon), i.e., systematic analysis
of $^8$B Coulomb dissociation for wide energy regions can be done.

\section{Summary and Conclusions}
\label{summary}

In the present paper we propose systematic analysis of
$^8$B Coulomb dissociation with the Asymptotic Normalization
Coefficient (ANC) method. An important advantage of the use of
the ANC method is that one can extract the astrophysical
factor $S_{17}(0)$ evaluating its uncertainties quantitatively,
in contrast to the previous analyses with the Virtual Photon
Theory (VPT).

In order to make accurate analysis of the measured spectra
in dissociation experiments, we use the method of Continuum-Discretized
Coupled-Channels (CDCC), which was developed by Kyushu group.
The CDCC + ANC analysis was found to work very well for
$^{58}$Ni($^8$B,$^7$Be$+p$)$^{58}$Ni at 25.8 MeV measured at
Notre Dame, and we obtained $S_{17}(0)=22.83\pm 0.51({\rm theo})\pm
2.28({\rm expt})$ ({\rm eVb}), which is consistent with both the latest
recommended value $19^{+4}_{-2}$ eVb~\cite{Adelberger}
and recent results of direct
measurements~\cite{Junghans,Baby}.

The CDCC + ANC analysis is expected to work well also at intermediate
energies. From a practical point of view, however, CDCC calculation
for $^8$B Coulomb dissociation at several tens of MeV/nucleon requires
extremely large modelspace, typically about 15,000 partial waves
are needed. In order to make efficient analysis at intermediate
energies, we introduce a new version of CDCC, that is,
the Eikonal-CDCC method (E-CDCC).
E-CDCC describes the center-of-mass (c.m.) motion between the projectile
and the target nucleus by a straight-line and
treats the excitation of the projectile explicitly, by constructing
discretized-continuum-states same as in CDCC.
The resultant E-CDCC equations are easily and safely be solved, since
they have a first-order differential form and no huge angular momenta,
in contrast to the CDCC equations.
Inclusion of the Coulomb distortion can be done with the use of
Coulomb wave functions instead of plane waves in eikonal (EK)
approximation.

One of the most important features of E-CDCC is that
the resultant scattering amplitude $f^{\rm E}$
has a very similar form to the quantum-mechanical (QM) one, i.e.,
$f^{\rm E}$ is expressed by the sum of partial amplitudes $f_L^{\rm E}$,
using relation between the angular momentum $L$ and the
impact parameter $b$.
This allows one to construct the hybrid (HY) amplitude $f^{\rm H}$ rather
straightforwardly;
$f^{\rm H}$ is given by the sum of the partial amplitudes
calculated by CDCC for smaller $L$, $\sum_{L\le L_{\rm C}} f_L^{\rm Q}$,
and those by E-CDCC for larger $L$, $\sum_{L > L_{\rm C}} f_L^{\rm E}$,
where $L_{\rm C}$ is the connecting value.
The HY calculation make the CDCC + ANC analysis much simple,
retaining its accuracy; all QM effects necessary can be included
through $f_L^{\rm Q}$.

The validity of the HY calculation is tested for
$^{58}$Ni($^8$B,$^7$Be$+p$)$^{58}$Ni at 240 MeV.
The HY calculation turned out to {\lq\lq}perfectly'' reproduce
the elastic and total breakup cross sections obtained by the QM
one, namely, the error is only less than 1\%;
the appropriate value of $L_{\rm C}$ is found to be 400 (200) for
the total-breakup (elastic) cross section, which is much smaller
than the required maximum value of $L$, i.e., $L_{\rm max}=4000$.
Calculation with a HY cross section, not HY amplitude, was found
to fail to reproduce the corresponding QM result, which shows
the importance of the interference between the
lower (QM) and higher (EK) regions of $L$.

In conclusion, systematic analysis of $^8$B Coulomb dissociation
with the ANC method and the HY Coupled-Channels (CC) calculation
is expected to accurately determine $S_{17}(0)$,
with reliable evaluation of its uncertainties.
An extracted $S_{17}(0)$ from the RIKEN, MSU and GSI data,
combined with that from the Notre Dame experiment,
will be reported in near future.

\section*{Acknowledgement}

The authors wish to thank M. Kawai, T. Motobayashi and T. Kajino
for fruitful discussions and encouragement.
We are indebted to the aid of JAERI and RCNP, Osaka University
for computation.
This work has been supported in part by the Grants-in-Aid for
Scientific Research
of the Ministry of Education, Science, Sports, and Culture of Japan
(Grant Nos.~14540271 and 12047233).

\end{document}